\documentclass[10pt,sn-nature]{sn-jnl}
\usepackage{amsmath, amssymb, amsfonts, stmaryrd, wasysym, graphicx, multirow, color, textcomp, subfigure}
\usepackage{manyfoot}
\usepackage{float}
\usepackage{color}
\usepackage{soul}
\usepackage{bm}
\usepackage{url}
\usepackage{booktabs}
\usepackage[normalem]{ulem}
\usepackage[dvipsnames]{xcolor}
\usepackage{geometry}
\setcitestyle{super,open={},close={}}

\begin{document}

\title{Emergence of stable meron quartets in twisted magnets}

\author*[1]{\fnm{Kyoung-Min}, \sur{Kim}} \email{kmkim@ibs.re.kr}
\author[2]{\fnm{Gyungchoon}, \sur{Go}}
\author[3]{\fnm{Moon Jip}, \sur{Park}}
\author*[2]{\fnm{Se Kwon}, \sur{Kim}} \email{sekwonkim@kaist.ac.kr}

\affil*[1]{ \orgdiv{Center for Theoretical Physics of Complex Systems}, \orgname{Institute for Basic Science}, \orgaddress{ \street{55 Expo-ro}, \city{Yuseong-gu}, \postcode{34126}, \state{Daejeon}, \country{Republic of Korea}}}

\affil*[2]{ \orgdiv{Department of Physics}, \orgname{Korea Advanced Institute of Science and Technology}, \orgaddress{ \street{291 Daehak-ro}, \city{Yuseong-gu}, \postcode{34141}, \state{Daejeon}, \country{Republic of Korea}}}

\affil[3]{ \orgdiv{Department of Physics}, \orgname{Hanyang Univercity}, \orgaddress{ \street{222 Wangsimni-ro}, \city{Seongdong-gu}, \postcode{04763}, \state{Seoul}, \country{Republic of Korea}}}

\abstract{The investigation of twist engineering in easy-axis magnetic systems has revealed the remarkable potential for generating topological spin textures, such as magnetic skyrmions. Here, by implementing twist engineering in easy-plane magnets, we introduce a novel approach to achieve fractional topological spin textures such as merons. Through atomistic spin simulations on twisted bilayer magnets, we demonstrate the formation of a stable double meron pair in two magnetic layers, which we refer to as the “Meron Quartet” (MQ). Unlike merons in a single pair, which is unstable against pair annihilation, the merons within the MQ exhibit exceptional stability against pair annihilation due to the protective localization mechanism induced by the twist that prevents the collision of the meron cores. Furthermore, we showcase that the stability of the MQ can be enhanced by adjusting the twist angle, resulting in increased resistance to external perturbations such as external magnetic fields. Our findings highlight the twisted magnet as a promising platform for investigating the intriguing properties of merons, enabling their realization as stable magnetic quasiparticles in van der Waals magnets.}


\maketitle
\tableofcontents
\listoffigures
\clearpage

\section{Introduction}

\begin{figure}[t!]
    \centering
    \includegraphics[width=.48\textwidth]{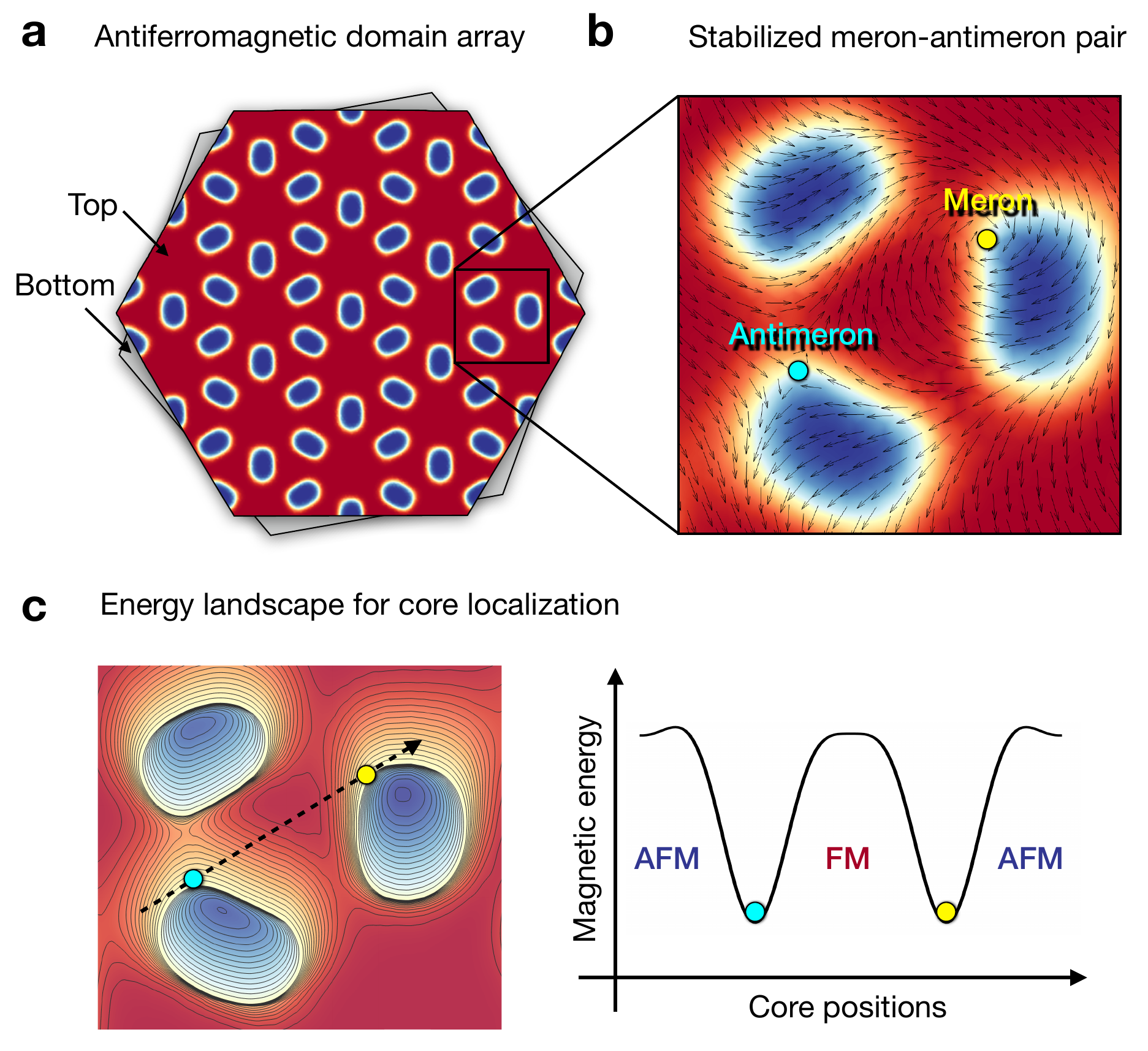}
    \caption[Schematic illustration]{\textbf{Schematic illustration of a stable meron-antimeron pair in a twisted magnet.} \textbf{a:} Twist-induced antiferromagnetic (AFM) domain array in a ferromagnetic (FM) order background. Red and blue colors indicate parallel and antiparallel spin alignments between the top and bottom layers, respectively. \textbf{b:} Emergence of a stable meron-antimeron pair. Arrows and circles depict their in-plane winding textures and core positions, respectively. \textbf{c:} Schematic energy landscape for localizing the cores within different AFM domains (left), and its schematic profile along the dashed line (right).}
    \label{fig1}
\end{figure}

The search for novel spin configurations motivated by fundamental interest and technological applications has led to the discovery of intriguing textures with nontrivial topology in various magnetic systems \cite{Nagaosa2013}. Specifically, vortex-type topological spin textures, so-called merons, have been observed in confined magnetic disks \cite{doi:10.1126/science.289.5481.930, PhysRevLett.108.067205, PhysRevLett.110.177201, PhysRevB.94.014433, 10.1063/1.1483386} and continuous thin films \cite{doi:10.1126/sciadv.aat3077, VanWaeyenberge2006, Ruotolo2009, PhysRevB.79.060407, Chmiel2018, Yu2018, Gao2019}. Recently, monolayer chromium trichloride (CrCl\textsubscript{3}), a two-dimensional (2D) van der Waals (vdW) magnetic crystal, has emerged as a promising candidate for achieving merons in this novel atomically thin limit \cite{doi:10.1126/science.abd5146, Lu2020, Augustin2021}. The intrinsic easy-plane magnetic anisotropy in such a system offers a pathway to attain in-plane swirling spin textures for merons. However, these merons are inherently unstable against pair annihilation and possess only a limited lifespan \cite{Augustin2021}. The mechanism responsible for stabilizing such merons has yet to be definitively determined, hampering future exploration of merons in 2D vdW magnets.

The field of twist engineering has opened up a fascinating realm of possibilities in generating topological spin textures in 2D vdW magnets. By harnessing moiré patterns, researchers have demonstrated the creation of skyrmion spin textures in lattice-mismatched heterostructures \cite{Tong2018, PhysRevB.104.L100406, PhysRevB.103.L140406} and twisted homo-bilayer systems \cite{PhysRevResearch.3.013027, Akram2021, Ghader2022, Zheng2022, Kim2023, Fumega_2023}, with a primary focus on Ising-type easy-axis magnetic systems. However, extending this approach to XY-type easy-plane magnets holds tremendous intrigue, as it enables exploration of captivating phenomena such as the Berezinskii-Kosterlitz-Thouless transition \cite{Berezinsky:1970fr, Kosterlitz:1973xp}, the emergence of merons \cite{Lu2020, Augustin2021}, and the potential discovery of hidden magnetic phases driven by strong spin fluctuations \cite{Burch2018}. The recent discovery of twisted magnets further enhances the allure, prompting a continued investigation into the captivating moiré effects in these systems \cite{Song2021, Xu2022, Xie2022, Xie2023}.

In this study, we investigate twist engineering in vdW magnets as a promising avenue to realize stable merons. By conducting atomistic spin simulations, we demonstrate that antiferromagnetic domain arrays in the twisted magnet \cite{Hejazi10721, Zheng2022, Kim2023, Song2021, Xu2022, Xie2022, Xie2023, Kim2023_trilayer} can be utilized to localize the cores of merons along the boundaries of their respective domains (Fig. \ref{fig1}\textbf{a-b}). This localization mechanism effectively preserves the stable spin configuration of the meron pair by separating their cores (Fig. \ref{fig1}\textbf{c}). Furthermore, we show that the stability of merons can be tuned by adjusting the twist angle, providing controllable resistance against external magnetic fields. These findings present a promising avenue for achieving stable merons as magnetic quasiparticles in vdW magnets, offering the opportunity to explore their captivating properties with significant flexibility through external stimuli \cite{Wang2018, Jiang2018} or the creation of heterostructures \cite{Gibertini2019}.

\section{Results}

\subsection{Antiferromagnetic domain array} \label{sec:MD}

\begin{figure*}[ht!]
    \centering
    \includegraphics[width=.9\textwidth]{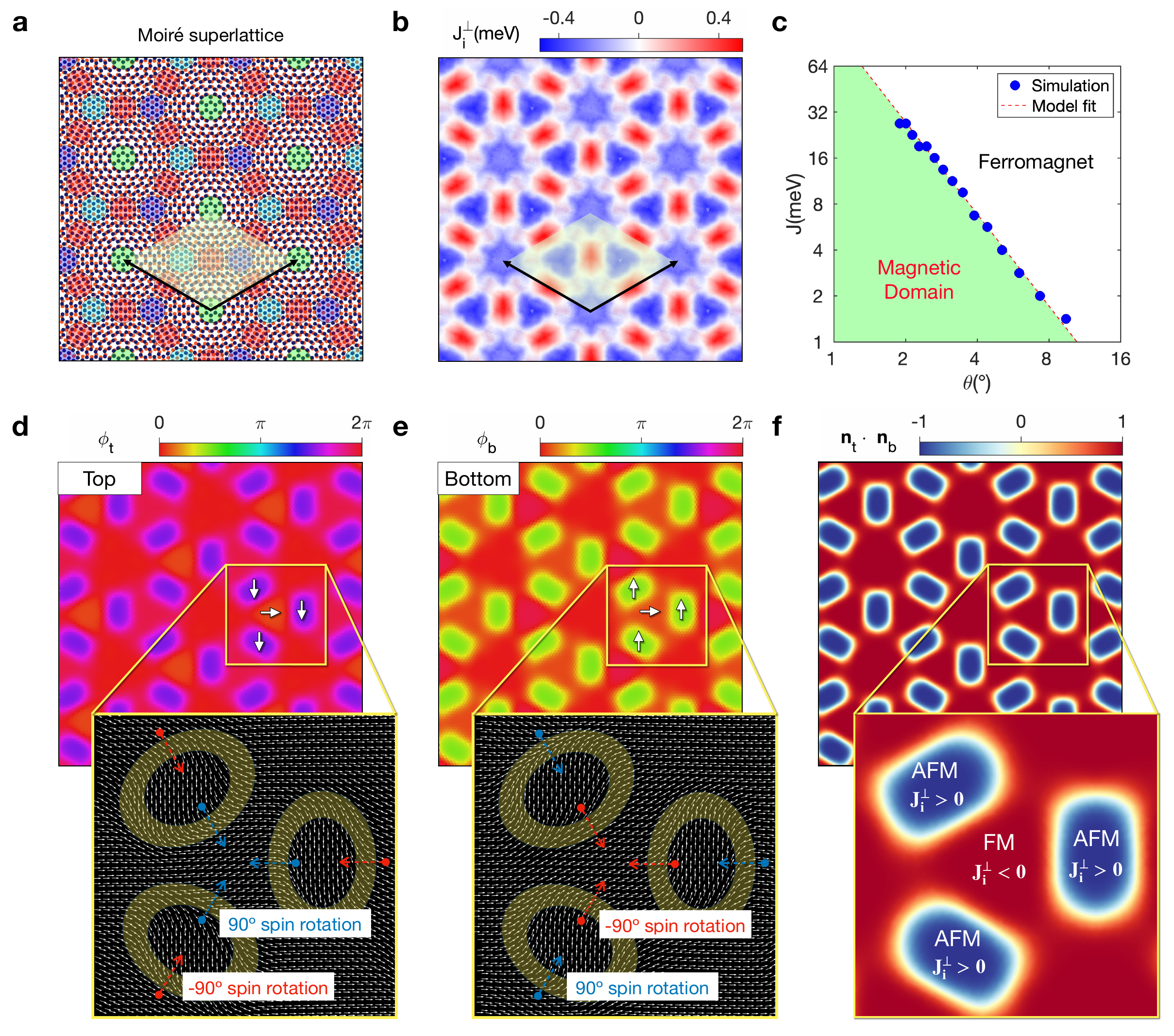}
    \caption[Emergence of AFM domain array]{\textbf{Emergence of AFM domain array.} \textbf{a:} Moiré superlattice for a twist angle $\theta=5.08$\textdegree{}. The colored circles denote local stacking patterns, including AA (green), AB (blue), BA (cyan), and monoclinic (red). The yellow rhombus and black arrows denote the unit cell and lattice vectors of the moiré superlattice, respectively. \textbf{b:} Modulation of the local interlayer exchange energy ($J_i^\perp$) computed for $\theta=1.02$\textdegree{}, with blue (red) color representing FM (AFM) coupling. \textbf{c:} Zero temperature magnetic phase diagram depicting FM phase (white) and magnetic domain phase (green), as a function of twist angle ($\theta$) and intralayer exchange ($J$). The markers represent the phase boundary obtained from numerical simulations, while the dotted line indicates the fitting curve computed from an effective continuum model (see Methods \ref{sec:theta_c_FM-MD}). \textbf{d-f:} Ground-state spin configuration in the magnetic domain phase for $J=2$ meV and $\theta=1.61$\textdegree{}. \textbf{d-e:} The color scales denote the phase angles ($\phi_{t,b}$) of the normalized spin vectors $\mathbf{n}_{t,b}=(\cos{\phi_{t,b}},\sin{\phi_{t,b}},0)$ in the top (\textbf{d}) and the bottom (\textbf{e}) layers, respectively. In the magnified images, the arrows denote the direction of $\mathbf{n}_{t}$ and $\mathbf{n}_{b}$, and the yellow areas highlight the domain walls with $90$\textdegree{} or $-90$\textdegree{} spin rotations. Here, ``t" and ``b" represent the top and bottom layers, respectively. \textbf{f:} The color scale denotes the relative orientation of the spin vectors between the two layers ($\mathbf{n}_t\cdot\mathbf{n}_b$), where red (blue) represents parallel (antiparallel) alignment. The magnified image highlights the correspondence between the interlayer exchange and  magnetic domain structure. }
    \label{fig2}
\end{figure*}

We construct twisted bilayer magnets by rotating two magnetic layers in a honeycomb lattice with a relative twist angle (Fig. \ref{fig2}\textbf{a}). These twisted magnets can be effectively described using a Heisenberg spin model given by \cite{Kim2023}:
\begin{align} \label{eq:spinH}
    H = & -\frac{J}{2}\sum_{l=t,b} \sum_{\langle i, j\rangle} \mathbf{S}_{i}^l \cdot \mathbf{S}_{j}^l + A \sum_{l=t,b} \sum_{i} \big(\bm{S}_{i}^l\cdot\hat{z}\big)^2 \nonumber \\
    & + \sum_{i, j} J_{ij}^\perp \mathbf{S}_{i}^t \cdot \mathbf{S}_{j}^b.
\end{align}
Here, $\mathbf{S}_{i}^{l}$ represents the spin at site $i$ on the top layer ($l=t$) and the bottom layer ($l=b$). $J$ represents the intralayer FM exchange interactions between nearest-neighbor spins. $A=0.1$ meV represents the single-ion anisotropy energy favoring in-plane magnetization. $J_{ij}^\perp$ represent the interlayer exchange interactions, which are adopted from previous ab-initio calculations on bilayer CrI\textsubscript{3} \cite{Kim2023}. Due to so-called stacking-dependent interlayer magnetism \cite{Sivadas2018, PhysRevMaterials.3.031001, Song2019, Li2019, doi:10.1126/science.aav1937, Akram2021}, $J_{ij}^\perp$ switch from FM to AFM coupling depending on the local stacking pattern between the two magnetic layers (Methods \ref{sec:J_perp}). Consequently, the interlayer exchange coupling exhibits the coexistence of AFM and FM interactions in the moiré superlattice accommodating various local stacking patterns (Fig. \ref{fig2}\textbf{a})\cite{PhysRevResearch.3.013027, Akram2021, Ghader2022, Zheng2022, Kim2023, Fumega_2023, Kim2023_trilayer}. We illustrate this behavior in Fig. \ref{fig2}\textbf{b} through the map of the local interlayer exchange energy $J_i^\perp = \sum_{j} J_{ij}^\perp$ computed in an FM configuration $\mathbf{S}_i^l=S\hat{z}$ \cite{PhysRevResearch.3.013027, Ghader2022, Kim2023, Kim2023_trilayer}. Specifically, in the monoclinic stacking region (red patches), $J_{i}^\perp$ exhibits AFM character ($J_{i}^\perp > 0$), indicating a tendency for the spins in the top and bottom layers to align antiparallel to each other. Conversely, in the other stacking regions, $J_{i}^\perp$ exhibits FM character ($J_{i}^\perp < 0$), signifying a preference for parallel alignment. As a result, the twisted magnet embeds local AFM patches in a background of FM coupling (Fig. \ref{fig2}\textbf{b}).

In this work, we investigate the influence of the AFM patches on the stabilization of merons. We first identify the magnetic phases of twisted easy-plane magnets. Our atomistic simulations on Eq. \eqref{eq:spinH}, conducted using an iterative optimization method (Methods \ref{sec:iom}), reveal the zero-temperature magnetic phase diagram shown in Fig. \ref{fig2}\textbf{c}. Within this diagram, we observe two distinct magnetic phases: an FM phase and a magnetic domain (MD) phase. The FM phase exhibits a uniform spin configuration with parallel alignment between the spins of the top and bottom layers. On the other hand, the MD phase exhibits antiparallel alignment within the AFM patches, while maintaining parallel alignment outside these patches to minimize the interlayer exchange energy (Fig. \ref{fig2}\textbf{d-f}). This contrasting AFM-FM order results in the formation of AFM domains within each AFM patch as well as domain walls surrounding the domains, characterized by spin rotations of $90$\textdegree{} and $-90$\textdegree{}. Furthermore, the AFM domains are arranged into an array structure across the superlattice, resembling a Kagome lattice. We dub this distinctive magnetic structure an AFM domain array.

We attribute the emergence of the AFM domain array to the amplified effect of interlayer exchange in the small twist angle regime ($\theta < \theta_{c1} \sim \sqrt{\bar{J}_\perp/J}$). In this regime, the formation of an AFM domain becomes energetically favorable as the reduction in the interlayer exchange energy ($\Delta E_\textrm{inter} \sim -\bar{J}_\perp\frac{L^2}{a^2}\sim -\bar{J}_\perp \theta^{-2}$) outweighs the increase in the intralayer exchange energy ($\Delta E_\textrm{intra}\sim J$). Here, we consider $\bar{J}_\perp$ as the average value of $J^\perp_i$ within the AFM patch. The patch is approximated as a disk with a radius denoted by $L\sim a/\theta$, where $a$ represents the lattice constant of the honeycomb lattice. Despite interlayer exchange being weaker than intralayer exchange, its effect is significantly amplified by the large size of the AFM patch ($\pi L^2$). Consequently, the formation of the AFM domain array is expected when the twist angle is sufficiently small. This phenomenon manifests in the phase diagram, which exhibits a consistent relationship $\theta_{c1}\sim\sqrt{\bar{J}_\perp/J}$ for the phase boundary between the FM and MD phases (dashed line in Fig. \ref{fig2}\textbf{c}).

\subsection{Emergence of stable merons: meron quartets} \label{sec:MQ}

\begin{figure*}[ht!]
    \centering
    \includegraphics[width=.9\textwidth]{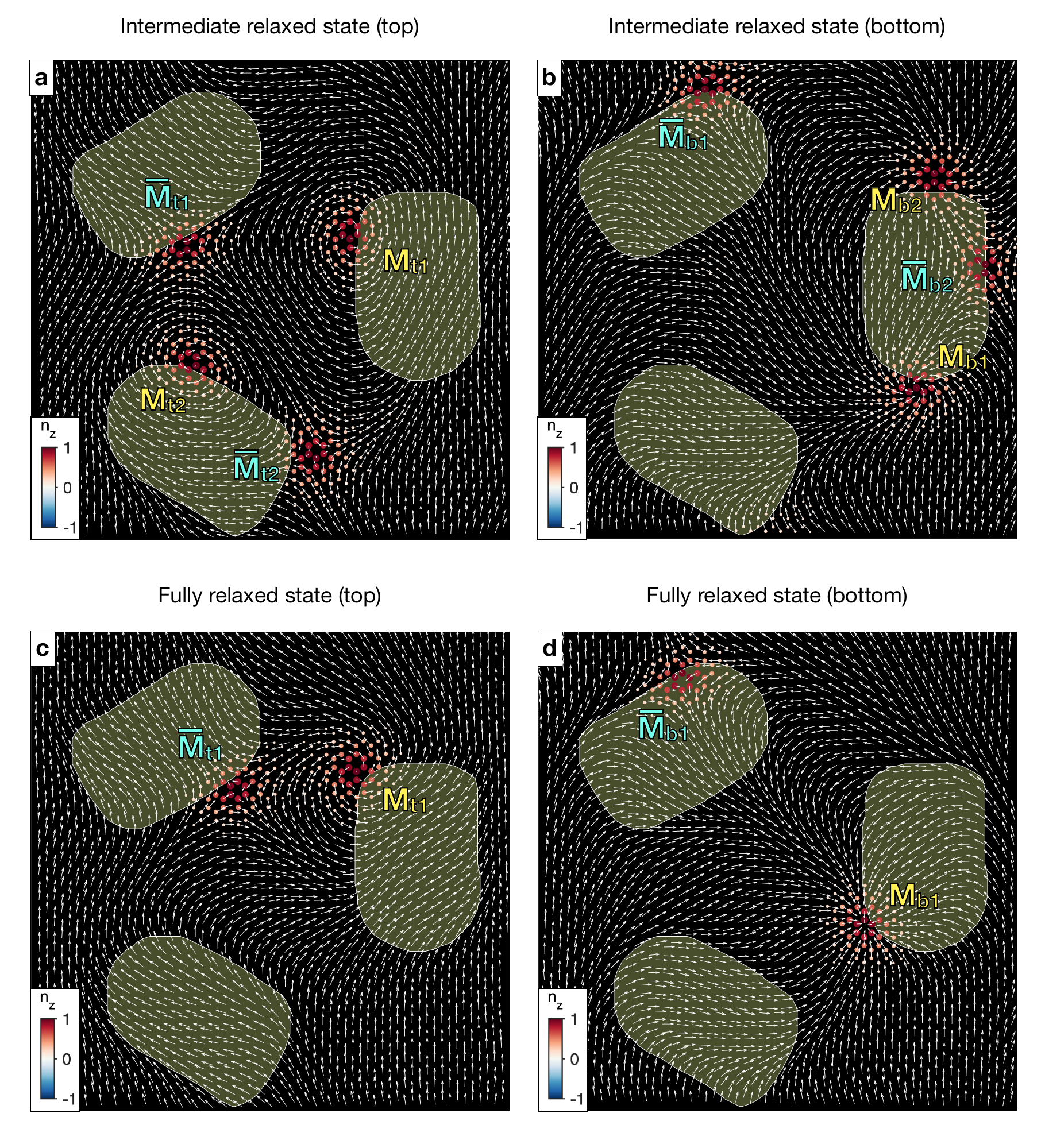}
    \caption[Emergence of stable meron-antimeron pairs]{\textbf{Snapshots depicting the relaxation process and the formation of stable meron-antimeron pairs.} \textbf{a-b:} Intermediate state illustrating the spontaneous generation of merons ($\textrm{M}_\textrm{t1}$, $\textrm{M}_\textrm{t2}$, $\textrm{M}_\textrm{b1}$, and $\textrm{M}_\textrm{b2}$) and antimerons ($\overline{\textrm{M}}_\textrm{t1}$, $\overline{\textrm{M}}_\textrm{t2}$, $\overline{\textrm{M}}_\textrm{b1}$, and $\overline{\textrm{M}}_\textrm{b2}$). \textbf{c-d:} Fully-relaxed state displaying stabilized meron-antimeron pairs ($\textrm{M}_\textrm{t1}$--$\textrm{M}_\textrm{t1}$, $\textrm{M}_\textrm{b1}$--$\textrm{M}_\textrm{b1}$), with the annihilation of pairs $\textrm{M}_\textrm{t2}$--$\overline{\textrm{M}}_\textrm{t2}$ and $\textrm{M}_\textrm{b2}$--$\overline{\textrm{M}}_\textrm{b2}$. Panels \textbf{a,c} and \textbf{b,d} represent the top and bottom layers, respectively, corresponding to a magnified area shown in Fig. \ref{fig2}\textbf{d-e}. Arrows indicate in-plane components, while the color scale in the markers represents out-of-plane components. Marker sizes are adjusted for better visibility. Shaded areas indicate AFM patches. The parameters $J=2$ meV and $\theta=1.61$\textdegree{} are utilized.}
    \label{fig3}
\end{figure*}

A meron is a vortex-like topological spin texture, characterized by an integer winding number. In contrast to conventional magnetic vortices, the core of a meron exhibits an out-of-plane polarization, resulting in a unique half-skyrmion number denoted as \cite{Yu2018}:
\begin{align} \label{eq:skyrmion_number}
    Q = \frac{1}{4\pi}\int dx\int dy(\partial_x\mathbf{n}\times\partial_y\mathbf{n})\cdot \mathbf{n} = p\cdot w = -\frac{1}{2}.
\end{align}
Here, the vector field $\mathbf{n}=(\sin{\vartheta}\cos{\varphi}, \sin{\vartheta}\sin{\varphi}, \cos{\vartheta})$ represents the orientation of spins. The polarity $p$ and vorticity $w$ of $\mathbf{n}$ are defined by $p=\frac{1}{2}[\cos\vartheta(r=\infty)-\cos\vartheta(r=0)]$ and $w=\frac{1}{2\pi}\oint_\gamma d\mathbf{l}\cdot\nabla\varphi$, respectively, where $\gamma$ is any contour that encircles the core. Merons possess two distinct characteristics, corresponding to the combinations $(w,p)=\left\{(+1, -\frac{1}{2}), (-1, +\frac{1}{2})\right\}$. Antimerons are counterparts to merons, possessing an opposing skyrmion number of $Q = +\frac{1}{2}$ with two distinct characteristics $(w,p)=\left\{(-1, -\frac{1}{2}), (+1, +\frac{1}{2})\right\}$.

In continuous magnetic systems, merons and antimerons typically exist as pairs with opposite winding numbers ($w=+1$ and $w=-1$) \cite{doi:10.1126/sciadv.aat3077, VanWaeyenberge2006, Ruotolo2009, PhysRevB.79.060407, Chmiel2018, Yu2018, Gao2019}. The formation of such pairs allows their swirling spin textures to cancel out away from the cores, resulting in localized spin configurations with finite energy. However, the mutual attraction between the cores renders the pairs inherently unstable, leading to pair annihilation during magnetization dynamics \cite{Hubert1998}. Consequently, in conventional untwisted magnetic systems, these magnetic textures are usually observed as transient states with a limited lifespan \cite{doi:10.1126/sciadv.aat3077, VanWaeyenberge2006, Gao2019, Augustin2021}.

In this work, we discover that the merons and antimerons in twisted magnets can evade pair annihilation by forming a double meron pair in two magnetic layers. To illustrate the emergence of such stable merons, we present a typical relaxation process of the magnetic state in Fig. \ref{fig3}, which is obtained through the relaxation of a random initial configuration (Methods \ref{sec:MD-MQ boundary}). In the intermediate magnetic state (Fig. \ref{fig3}\textbf{a-b}), we observe the spontaneous formation of four merons ($\textrm{M}_\textrm{t1}$, $\textrm{M}_\textrm{t2}$, $\textrm{M}_\textrm{b1}$, and $\textrm{M}_\textrm{b2}$) and four antimerons ($\overline{\textrm{M}}_\textrm{t1}$, $\overline{\textrm{M}}_\textrm{t2}$, $\overline{\textrm{M}}_\textrm{b1}$, and $\overline{\textrm{M}}_\textrm{b2}$) on the top and bottom layers. Upon subsequent relaxation, the intra-patch pairs, i.e. the meron-antimeron pairs occupying the same AFM patch within the same layer such as $\textrm{M}_\textrm{t2}$--$\overline{\textrm{M}}_\textrm{t2}$ and $\textrm{M}_\textrm{b2}$--$\overline{\textrm{M}}_\textrm{b2}$, undergo pair annihilation due to the attractive interactions driven by the intralayer exchange interactions. However, the inter-patch pairs, i.e. the meron-antimeron pairs occupying different AFM patches within the same layer such as $\textrm{M}_\textrm{t1}$--$\overline{\textrm{M}}_\textrm{t1}$ and $\textrm{M}_\textrm{b1}$--$\overline{\textrm{M}}_\textrm{b1}$, remain robust against pair annihilation due to their mutual correlation facilitated by the interlayer coupling. As a result, the fully relaxed state (Fig. \ref{fig3}\textbf{c-d}) accommodates only this correlated double meron pair.

We attribute the stabilization of the double meron pair to the localization of their cores. During the relaxation process (Supplementary Video 1), we observe that the cores shift exclusively along the boundaries of the AFM patches. This behavior arises from the bulk energy minimization condition: the requirement to minimize the interlayer exchange energy over the bulk region by maintaining the AFM domain configuration, i.e. antiparallel and parallel alignments inside and outside the AFM patches as depicted in Fig. \ref{fig2}\textbf{f}. Enforcing such a condition localizes the cores within their respective AFM patches by constraining their motions along the patch boundaries and prohibiting them from transferring to other patches. This localization mechanism preserves the inter-patch pairs by ensuring the separation of their cores and protecting them against pair annihilation. However, the localization mechanism cannot preserve the intra-patch pairs, as their cores reside within the same patch (Supplementary Video 2).

\begin{figure*}[t!]
    \centering
    \includegraphics[width=\textwidth]{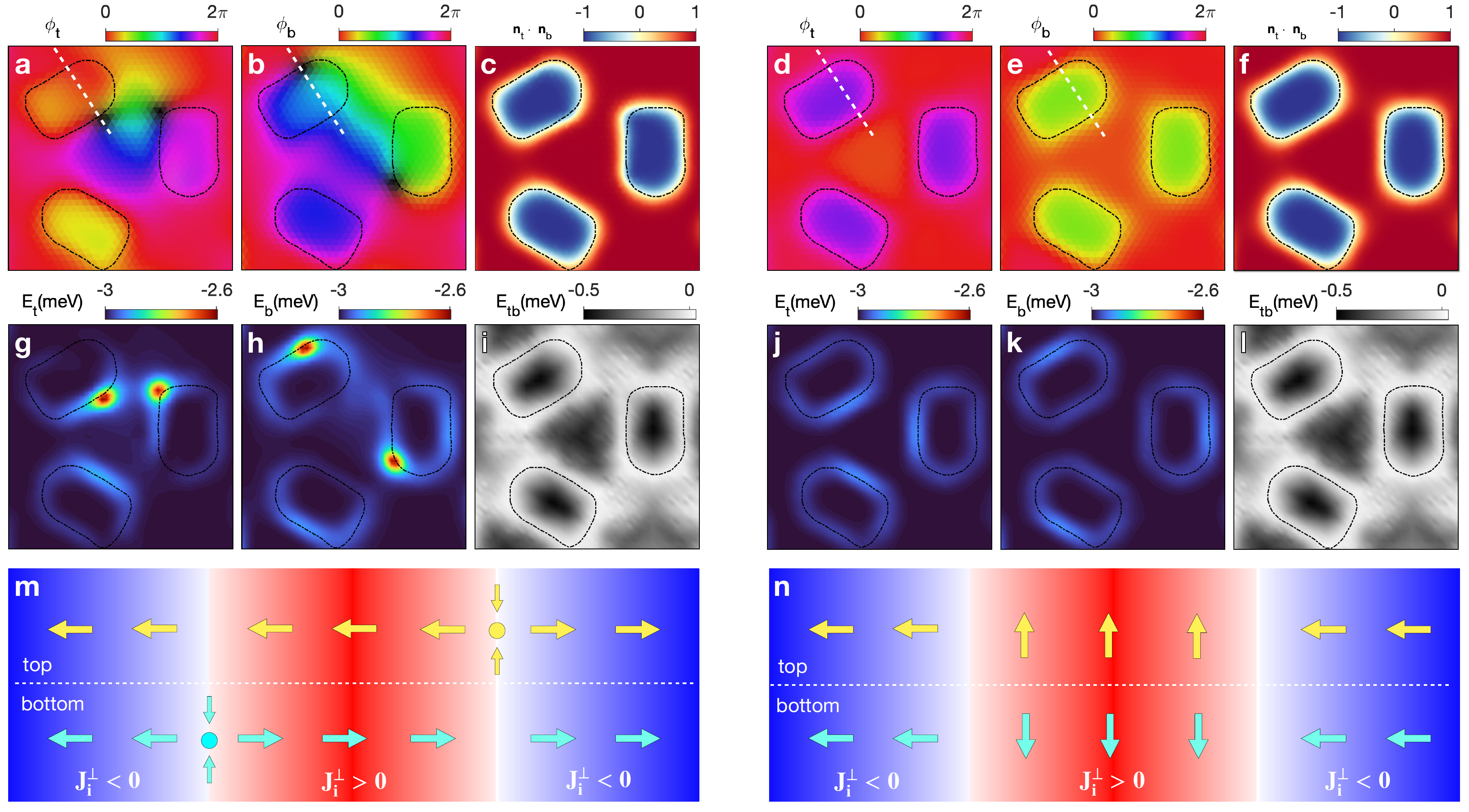}
    \caption[Comparison of MQ and MD]{\textbf{Comparison of meron quartet (MQ) state to magnetic domain (MD) state.} \textbf{a-c/d-f:} Spin configurations of the MQ (\textbf{a-c}) and MD (\textbf{d-f}) state, corresponding to Fig. \ref{fig3}\textbf{b,d} and Fig. \ref{fig2}\textbf{d-f}, respectively. \textbf{a-b/d-e:} Phase angles ($\phi_{t,b}$) of the normalized spin vectors ($\mathbf{n}_{t,b}=(\cos{\phi_{t,b}},\sin{\phi_{t,b}},0)$) in the top (\textbf{a/d}) and bottom (\textbf{b/e}) layers. In \textbf{a-b}, the black color indicates out-of-plane polarization. \textbf{c/f:} Relative orientation ($\mathbf{n}_t\cdot\mathbf{n}_b$) between the top and bottom layers, with red (blue) denoting parallel (antiparallel) alignment. \textbf{g-i/j-l:} Local magnetic energy maps corresponding to the spin configurations (\textbf{a-c/d-f}). \textbf{g-h/j-k:} Intralayer exchange energy plus single-ion anisotropy energy ($\textrm{E}_\textrm{t,b}$) in the top (\textbf{g/j}) and bottom (\textbf{h/k}) layers, respectively. \textbf{i/l:} Interlayer exchange energy ($\textrm{E}_\textrm{tb}$). In \textbf{a-i}, the dashed lines denote the boundaries of AFM patches. \textbf{m/n:} Schematic illustration of the MQ (\textbf{m}) and MD (\textbf{n}) states across a single AFM patch, corresponding to the dotted lines shown in \textbf{a-b} and \textbf{d-e}, respectively. Red and blue colors depict the AFM patch region ($J_i^\perp>0$) and FM coupling background ($J_i^\perp<0$), respectively. Yellow and blue arrows represent spin orientations in the top and bottom layers, respectively.} 
    \label{fig4}
\end{figure*}

The localization mechanism of the meron pair can be understood by considering the effective confining potential arising from the interlayer coupling (Fig. \ref{fig1}\textbf{c}). The cores of merons in one layer (e.g., the top layer) experience an effective potential generated by the other layer (e.g., the bottom layer) through the interlayer exchange coupling (Fig. \ref{fig4}\textbf{m}). This potential reaches its minimum energy along the boundaries of the AFM patches due to the bulk energy minimization condition. Any displacement of the cores away from these boundaries results in increased energy relative to the minimum, creating potential wells along the AFM patch boundaries. These wells act as confining forces, effectively localizing the cores within them. Furthermore, the establishment of such potential wells requires the presence of two counterpart merons in the bottom layer to facilitate the bulk energy minimization condition. Consequently, the creation of four merons is required to protect the merons via the confining potential.

Based on our findings, we introduce a novel magnetic state dubbed the “Meron Quartet” (MQ) state, which consists of four merons, two for each layer, as depicted in Fig. \ref{fig4}. This state exhibits two key characteristics: Firstly, each layer contains two vortices with opposite winding numbers ($w=+1$ and $w=-1$), ensuring the total winding number cancels out. Secondly, each occupied AFM patch harbors two vortices with the same winding numbers ($w=+1$ or $w=-1$) in both the top and bottom layers. These specific arrangements enable the MQ state to realize stable meron pairs through the implementation of the confining potential. Furthermore, the MQ state minimizes the interlayer exchange energy over the bulk region (Fig. \ref{fig4}\textbf{c,i}), as observed in the ground MD state (Fig. \ref{fig4}\textbf{f,l}). The distinction between the MQ and MD states lies solely in the localized core energy (Fig. \ref{fig4}\textbf{g-h} vs. \textbf{j-k}). As a result, the MQ state exhibits a low magnetic energy of -3.064 meV per spin, comparable to the energy of the ground MD state (-3.071 meV), despite its intricate spin textures arising from the presence of merons (Fig. \ref{fig4}\textbf{a-b}). In other words, the MQ state naturally accommodates the stable correlated pairs of merons, i.e. the meron quartet, while satisfying the bulk energy minimization condition (Fig. \ref{fig4}\textbf{m}), similar to the ground MD state (Fig. \ref{fig4}\textbf{n}).

\subsection{Stability of meron quartet states} \label{sec:stability}

\begin{figure*}
    \centering
    \includegraphics[width=\textwidth]{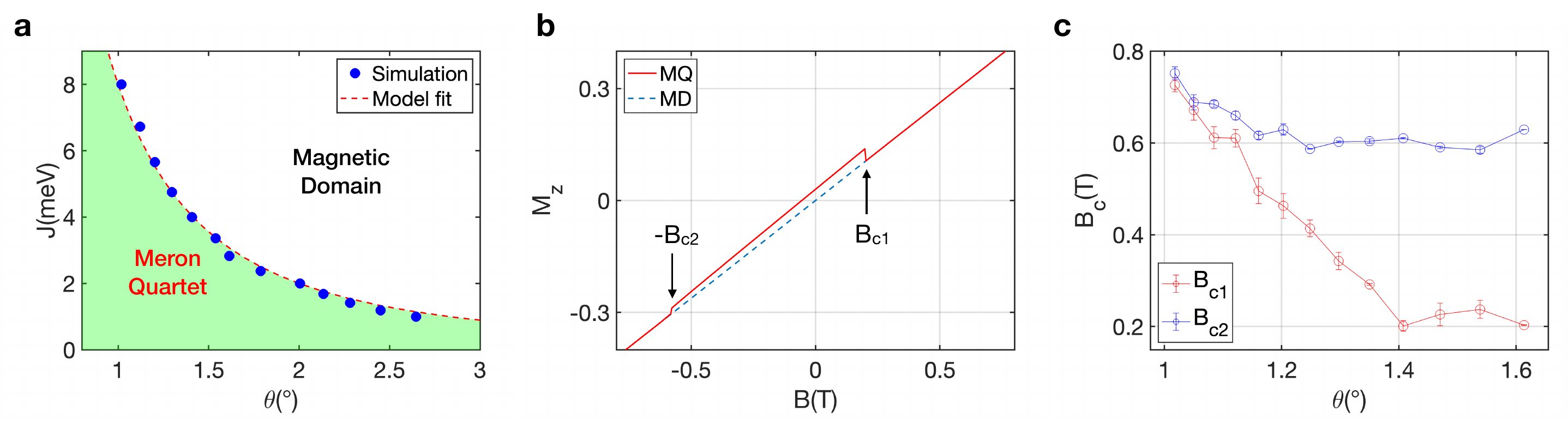}
    \caption[Stability of MQ]{\textbf{a:} Magnetic phase diagram illustrating the meron quartet phase (green) and the magnetic domain phase (white), as a function of twist angle ($\theta$) and intralayer exchange ($J$). The markers represent the phase boundary determined through numerical simulations, while the dashed line represents the phenomenological fitting curve $\theta_{c2} = \sqrt{J_{c2}/J}$, where the fitting parameter $J_{c2}$ is found to be $8$ meV. \textbf{b:} Out-of-plane net magnetization ($M_z=\frac{1}{N}\sum_{l=t,b}\sum_{i}n_{i,z}^l$) as a function of an external magnetic field in the out-of-plane direction ($B$), with the solid and dashed lines corresponding to the MQ and MD states, as depicted in Fig. \ref{fig4}\textbf{a-c} and \textbf{d-f}, respectively. The arrows mark the critical field strengths ($B_{c1}$ and $B_{c2}$) that signify the degradation of the MQ state to the MD state, as shown by the merging of the two curves at $B_{c1}$ and $-B_{c2}$. \textbf{c}: Evolution of $B_{c1}$ and $B_{c2}$ with twist angle ($\theta$). Error bars denote standard errors from different samples. The parameter $J=2$meV is utilized. }
    \label{fig5}
\end{figure*}

We observe that the MQ state acquires metastability in the small twist angle regime (green area in Fig. \ref{fig5}\textbf{a}). This phenomenon is attributed to the enhancement of the confining potential in such a regime. To elucidate this, we consider the transition of the MQ state to the MD state. This transition necessitates the mutual attractions of the meron cores. However, such attraction leads to an inevitable increase in the interlayer exchange energy, as it disrupts the bulk energy minimization condition, as illustrated in Fig. \ref{fig4}\textbf{c,i}. We estimate this energy increase as $\Delta E_\textrm{tb}\sim J_\perp^\textrm{FM}\frac{dL}{a^2}$, where $J_\perp^\textrm{FM}$ represents the FM interlayer exchange near an AFM patch, and $d$ signifies the displacement of a core from its equilibrium position due to the attraction. The term $\frac{dL}{a^2}\sim\frac{d}{a\theta}$ corresponds to the number of spins that undergo unfavorable ordering for the interlayer exchange, which increases as the twist angle $\theta$ decreases. Consequently, this energy cost escalates as the twist angle $\theta$ decreases, creating a substantial energy barrier for the transition to the MD state, which corresponds to the confining potential mentioned before.

This high energy barrier surpasses the attractive force between merons and antimerons, stabilizing the MQ state. The attractive force is mediated by the intralayer exchange interaction energy between a meron and an antimeron within the same layer. This energy is akin to the Coulomb energy between a vortex and an antivortex in the XY model, denoted as $E_\textrm{C}\sim J\ln{(R/a)}$ \cite{Hubert1998}, where $R\sim a/\theta$ represents the distance between the cores. The attraction of the cores by a displacement of $d$ can potentially reduce the Coulomb energy by $\Delta E_\textrm{C} \sim -J d/R\sim-Jd\theta/a$. However, this energy reduction $\Delta E_\textrm{C}$ is surpassed by the energy increase $\Delta E_\textrm{tb}$ in the small twist angle regime ($\theta<\theta_{c2}\sim \sqrt{J_\perp^\textrm{FM}/J}$). As a result, the attraction of the cores is effectively prohibited in such a regime, leading to the stabilization of the MQ state. We find that this phenomenon is evident in the phase diagram, which exhibits a consistent relationship $\theta_{c2} = \sqrt{J_{c2}/J}$ (dashed line in Fig. \ref{fig5}\textbf{a}), where $J_{c2}$ is a fitting parameter proportional to $J_\perp^\textrm{FM}$.

The high energy barrier also indicates the enhanced stability of the MQ state against external perturbations, such as external magnetic fields, in the small twist angle regime. To further illustrate this stability, we incorporate the Zeeman term:
\begin{align} \label{eq:zeeman}
    H_\textrm{Zeeman}=-g\mu_BB\sum_{l=t,b}\sum_{i}S_{i,z}^l,
\end{align}
where $B$ represents an external magnetic field applied in the out-of-plane direction. Through the systematic examination of the behavior of the MQ state under the influence of the Zeeman term (Methods \ref{sec:critical_field}), we identify the critical field strength for the destruction of the MQ state, as illustrated in Fig. \ref{fig5}\textbf{b}. The critical field strength differs depending on the relative orientation of the applied field with respect to the polarity of the meron and antimeron cores, with the antiparallel field exhibiting a much higher critical field strength ($B_{c2}$) compared to the parallel field ($B_{c1}$). Nevertheless, both critical field strengths are significantly enhanced as the twist angle decreases (Fig. \ref{fig5}\textbf{c}). This corroborates the enhanced stability of the MQ state in the small twist angle regime.

\subsection{Diverse forms of meron quartets} \label{sec:multiplicity}

We discover that the twisted magnet can realize stable merons in diverse forms with a different total skyrmion number ($Q_\textrm{tot}=0, \pm1, \pm2$) and distinct combinations of merons and antimerons between two magnetic layers in addition to the specific configuration shown in Fig. \ref{fig4}\textbf{a-c}. Due to the condition of the vanishing of the total winding number, two vortices consisting of each inter-patch pair must exhibit opposite winding numbers $w=+1$ and $w=-1$, respectively. However, the cores have the flexibility to possess their own polarity, which can be either $p=-\frac{1}{2}$ or $p=+\frac{1}{2}$. This gives rise to four potential configurations for each pair: (i) $\textrm{M}$--$\overline{\textrm{M}}$, (ii) $\overline{\textrm{M}}$--$\textrm{M}$, (iii) $\textrm{M}$--$\textrm{M}$, and (iv) $\overline{\textrm{M}}$--$\overline{\textrm{M}}$. Furthermore, these configurations can independently occur in the top and bottom layers, resulting in a total of sixteen potential configurations for the MQ state, with distinct numbers of merons and antimerons in each layer. Our investigation employing general random initial configurations confirms the emergence of such diverse configurations (Fig. \ref{figE5}). This observation highlights the flexibility in achieving merons and antimerons in the twisted magnet, setting it apart from conventional magnetic systems, which typically have fixed skyrmion numbers \cite{Nagaosa2013, Gao2019}.

\section{Discussion}

We have shown that the AFM domain array induced by the twist provides a favorable environment for hosting stable merons. Once these topological defects form, their destruction is impeded due to the constraints imposed by the bulk energy minimization condition. Moreover, our theory suggests the feasibility of realizing such stable merons and their enhanced robustness in the small twist angle regime.

We propose our theory can be applied to CrCl\textsubscript{3}, which exhibits two essential factors for hosting the meron quartet: easy-plane magnetic anisotropy \cite{Lu2020} and stacking-dependent interlayer magnetism \cite{Akram2021}. Notably, merons have been observed in monolayer CrCl\textsubscript{3} \cite{Lu2020, Augustin2021}. In this context, twist engineering offers an effective approach to realizing the meron quartet. Another potential application lies in CrI\textsubscript{3}, where the modification of its intrinsic easy-axis magnetic anisotropy to easy-plane anisotropy can be achieved through experimental control, such as gate-voltage tuning \cite{Tang2023}.

For experimental observations, we propose utilizing scanning magnetometry techniques with nitrogen-vacancy centers \cite{Song2021}, as well as Lorentz transmission electron microscopy \cite{Ding2020} and magnetic transmission soft X-ray microscopy \cite{Gao2019}, to directly observe meron pairs ranging in size from 80 to 20 nm at different twist angles ($\theta$ = 0.5\textdegree{} to 2\textdegree{}). Indirect measurements can involve detecting anomalous kinks in the magnetization curve, which can serve as an indication of the presence of merons. Techniques such as the magneto-optical Kerr effect, commonly employed in the study of 2D vdW magnets \cite{Xu2022, Xie2022}, can offer valuable insights for such indirect measurements.

Future research should consider incorporating various magnetic interactions present in vdW magnets that were overlooked in our current model, such as exchange anisotropy, interactions beyond nearest neighbors \cite{Augustin2021}, the Dzyaloshinskii-Moriya interaction, and magnetic dipole-dipole interactions \cite{Lu2020}. An important research question to explore is how these additional interactions impact the stabilization of the meron quartet and their behaviors in vdW magnets.

We highlight that the discovery and realization of merons in 2D vdW magnets via twist open a unique avenue for investigating their fascinating properties with remarkable flexibility, through either external stimuli \cite{Wang2018, Jiang2018} or the creation of heterostructures \cite{Gibertini2019}. This significant breakthrough not only deepens our understanding of these fractionalized topological spin textures in magnets but also holds great promise for future technological advancements.

\bibliography{ref.bib}

\clearpage

\section{Methods} \label{sec:method}

\subsection{Moiré superlattice} \label{sec:superlattice}

We utilized commensurate moiré superlattice structures generated by rotating one layer of an aligned honeycomb lattice bilayer relative to the other layer. The rotation is centered at one of the hexagonal centers and is defined by the angle $\theta=\arccos{\left(\frac{m^2+n^2+4mn}{2m^2+2n^2+2mn}\right)}$. Our specific focus was on the case of $n=m+1$, where the superlattice periodicity aligns with the periodicity of moiré patterns.

\subsection{Interlayer Heisenberg exchange interactions} \label{sec:J_perp}

\begin{figure*}[t!]
    \centering
    \includegraphics[width=\textwidth]{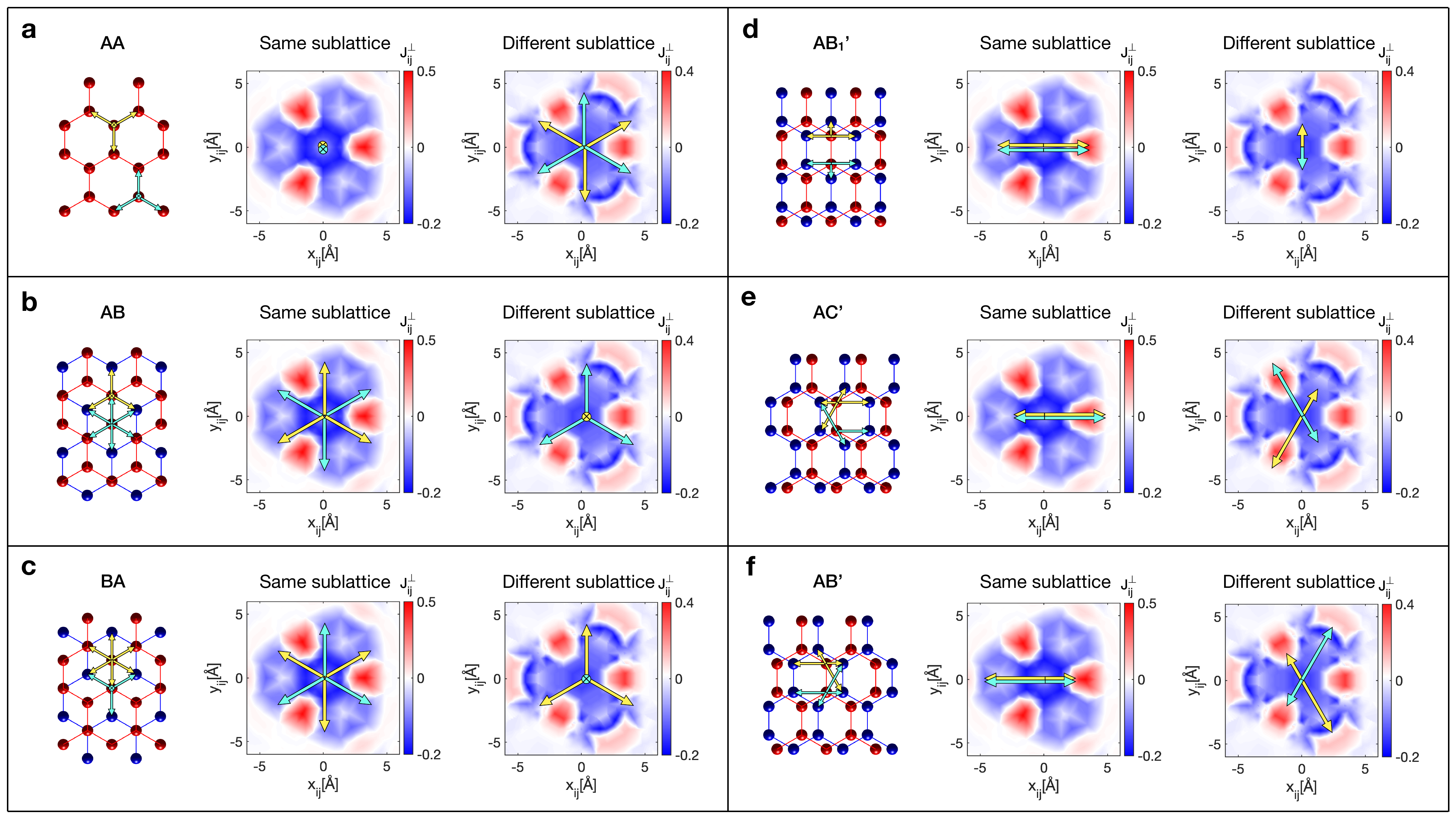}
   \caption[Ext. Data: Interlayer exchange interactions]{\textbf{Illustration of the procedure to determine interlayer Heisenberg exchange interactions.} \textbf{a-f:} Each panel depicts the procedure for determining interlayer Heisenberg exchange interactions $J_{ij}^\perp$ in different local stacking patterns. The left images show the local stacking patterns, with red and blue spheres representing the positions of local spins in the top and bottom layers, respectively. Yellow (blue) arrows indicate the parallel coordinate displacements from A (B) sublattice spins to their neighboring spins. The middle (right) color plot labeled ``Same sublattice" (``Different sublattice") depicts $J_{ij}^\perp$ as a function of the parallel coordinate displacements between two spins on the top and bottom layers $(x_{ij},y_{ij})=(x_i-x_j, y_i-y_j)$, where red and blue colors indicate AFM and FM characters, respectively. The same displacement vectors shown in the left images are drawn onto these maps, illustrating the interpolation process for determining $J_{ij}^\perp$ corresponding to each bond. The coupling maps were adopted from previous ab initio calculations on bilayer CrI\textsubscript{3} \cite{Kim2023}. }
    \label{figE1}
\end{figure*}

We determined $J_{ij}^\perp$ in our spin model by utilizing the coupling maps for interlayer Heisenberg exchange interactions derived from previous ab initio calculations on bilayer CrI\textsubscript{3} \cite{Kim2023}, as depicted in the color plots of Fig. \ref{figE1}. Our method \cite{Zheng2022, Kim2023} involves computing the parallel coordinate displacements $(x_{ij}, y_{ij})$ for all bonds between spins from the top and bottom layers, respectively. These displacements are then interpolated onto the coupling maps, which are functions of $(x_{ij}, y_{ij})$, to obtain the values of $J_{ij}^\perp$ for each bond. Fig. \ref{figE1}\textbf{a-f} illustrate this procedure for representative stacking patterns within the moiré superlattice. There are two distinct types of bonds: (i) bonds between two spins on the same sublattice and (ii) bonds between two spins on different sublattices. The coupling maps labeled ``Same sublattice" and ``Different sublattice" are utilized for the former and latter cases, respectively. The interactions for bond distances $\sqrt{x_{ij}^2+y_{ij}^2}$ larger than $\sqrt{3}a \approx 5.2$\AA{} are neglected, as their contribution is negligible. 

\subsection{Iterative optimization method} \label{sec:iom}

\begin{figure*}[t!]
    \centering
   \includegraphics[width=.98\textwidth]{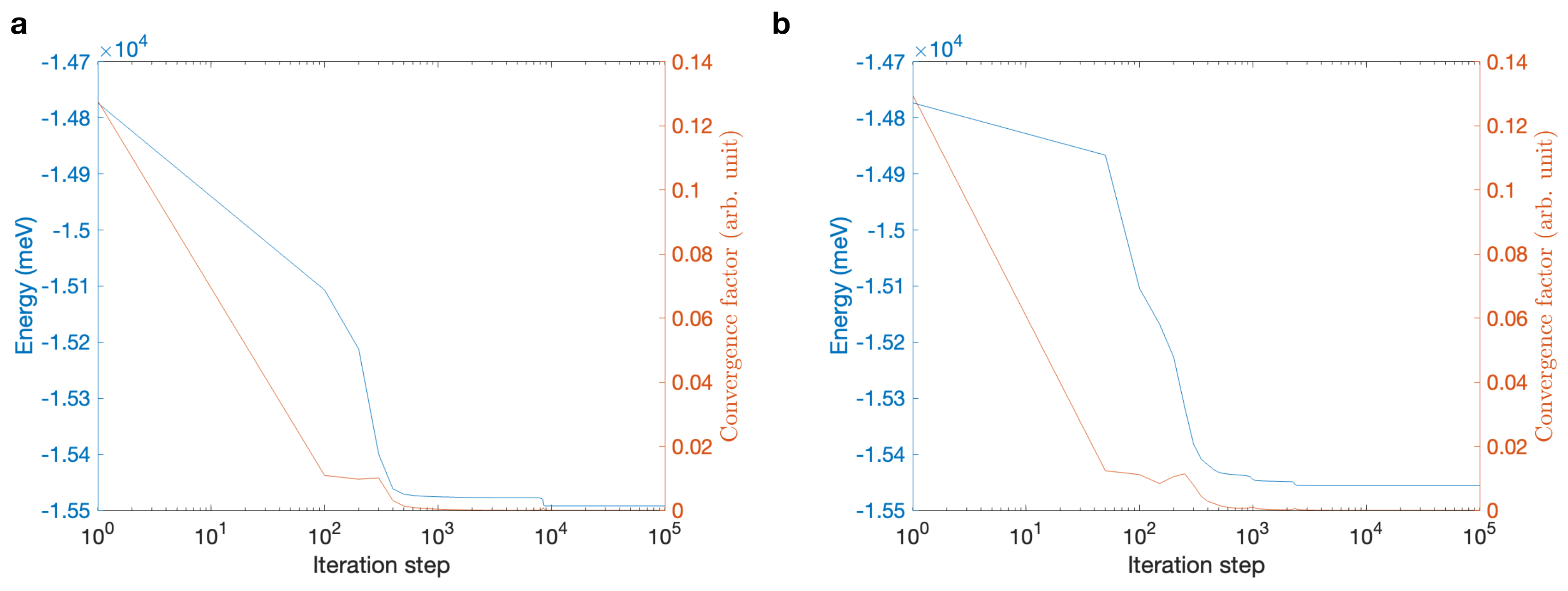}
    \caption[Ext. Data: Relaxation method]{Relaxation of spin configurations using the iterative optimization method. Panel \textbf{a} represents the MD state, as depicted in Fig. \ref{fig4}\textbf{d-f}. Panel \textbf{b} corresponds to the MQ state, as illustrated in Fig. \ref{fig4}\textbf{a-c}. The blue curves represent the total magnetic energy, and the red curves represent the convergence factor in the relaxation defined as $\frac{1}{N}\sum_{i}|\mathbf{h}_i^\perp|$.}
    \label{figE2}
\end{figure*}

We obtained a stable spin configuration using an iterative optimization method. In this procedure, we update a trial configuration by subtracting the normalized spin vector $\mathbf{n}_i = \mathbf{S}_i / |\mathbf{S}_i|$ with the gradient vector of the spin Hamiltonian $H$, denoted as $\mathbf{h}_i \equiv \frac{\partial H[\{\mathbf{n}_i\}]}{\partial \mathbf{n}_i}$. Here, $\mathbf{S}_i$ represents the classical spin vector at the $i$-th site. The update rule is given by:
\begin{equation}
    \mathbf{n}_i^{(s+1)}=\frac{\mathbf{n}_i^{(s)}-c \mathbf{h}_i^{(s)}}{\left|\mathbf{n}_i^{(s)}-c \mathbf{h}_i^{(s)}\right|},
\end{equation}
where $(s)$ denotes the iteration step, and $c$ controls the update rate. The iterative process continues until $\mathbf{h}_i^{(s)}$ becomes parallel to $\mathbf{n}_i^{(s)}$, implying that further updates will not change the spin configuration. At this stage, the magnetic energy is minimized. To ensure numerical convergence, we used the condition $\frac{1}{N}\sum_{i}|\mathbf{h}_i^\perp|<10^{-10}$, where $\mathbf{h}_i^\perp=\mathbf{h}_i-\mathbf{n}_i(\mathbf{n}_i\cdot\mathbf{h}_i)$. This convergence condition is identical to the stability condition in the Landau-Lifshitz-Gilbert equation describing the magnetic dynamics. We incorporated periodic boundary conditions during the iterative updates to maintain consistency across the boundaries of the moiré unit cell.

\subsection{Local magnetic energy maps} \label{sec:energy_map}

We computed the local energy maps in Fig. \ref{fig4} and Supplementary Videos (\ref{sec:movie1}, \ref{sec:movie2}) by utilizing the following equations:
\begin{subequations} \label{eq:localE}
\begin{align}
    E_\textrm{tot}(\mathbf{r}_i) & = E_\textrm{t}(\mathbf{r}_i) + E_\textrm{b}(\mathbf{r}_i) + E_\textrm{tb}(\mathbf{r}_i),  \label{eq:E_tot} \\
    E_\textrm{t}(\mathbf{r}_i) & = -\frac{JS^2}{2}\sum_{j\in\textrm{N.N}(i)} \mathbf{n}_{i}^{t} \cdot \mathbf{n}_{j}^{t} + AS^2  \left(\mathbf{n}_{i}^{t} \cdot \hat{z}\right)^2,  \label{eq:Et} \\
    E_\textrm{b}(\mathbf{r}_i) & = -\frac{JS^2}{2}\sum_{j\in\textrm{N.N}(i)} \mathbf{n}_{i}^{b} \cdot \mathbf{n}_{j}^{b} + AS^2  \left(\mathbf{n}_{i}^{b} \cdot \hat{z}\right)^2,  \label{eq:Eb} \\
    E_\textrm{tb}(\mathbf{r}_i) & = S^2\sum_{j} J_{ij}^\perp \mathbf{n}_{i}^t \cdot \mathbf{n}_{j}^b. \label{eq:Etb}
\end{align}
\end{subequations}
Here, $\mathbf{r}_i$ represents the position of $\mathbf{n}_i$, and $\text{N.N}(i)$ denotes the nearest neighbors of the $i$-th site. We have set $S=1$ for simplicity.

\subsection{Determination of FM-MD phase boundary} \label{sec:FM-MD boundary}

\begin{figure*}[t!]
    \centering
   \includegraphics[width=.66\textwidth]{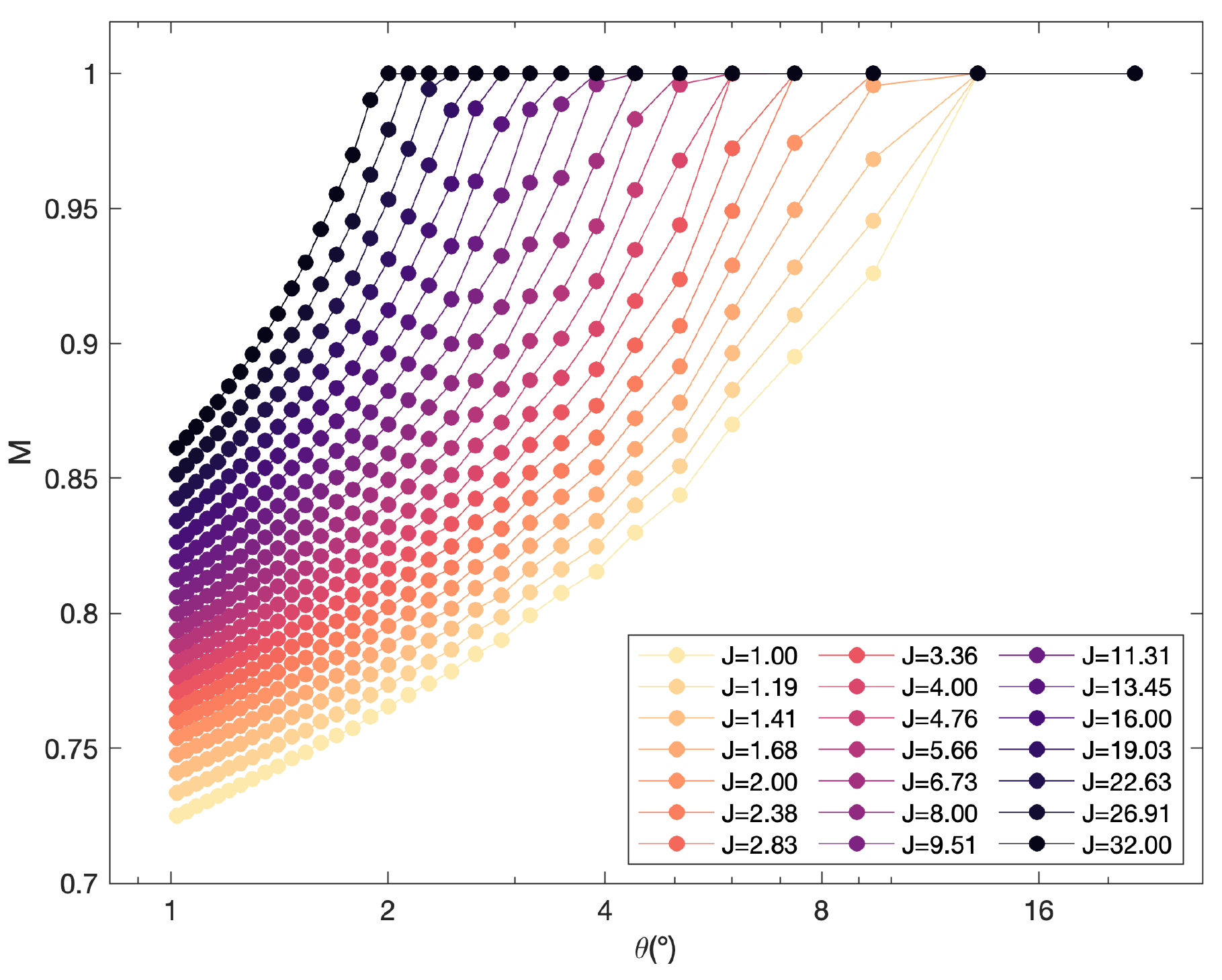}
    \caption[Ext. Data: FM-MD transition]{Ferromagnetic order parameter ($M$), defined by Eq. \eqref{eq:M}, plotted as a function of twist angle ($\theta$) for various intralayer exchange values ($J$) (in meV unit).}
\label{figE3}
\end{figure*}

We utilized random initial configurations represented as $\{\mathbf{n}_i^{(0)}\} = \left\{(u_i,v_i,w_i) \right\}$, where $u_i$, $v_i$, and $w_i$ are random numbers within the range of $[-1,1]$. Normalization was performed to ensure unit length for all $\mathbf{n}_i^{(0)}$. These initial configurations were relaxed through the iterative optimization method to obtain the ground state configuration $\{\mathbf{n}_i\}$ for each $\theta$ and $J$. From these ground states, we computed the ferromagnetic order parameter defined as: 
\begin{align} \label{eq:M}
    M = \frac{1}{N}\left|\sum_i \mathbf{n}_i \right|.
\end{align}
The results are shown in Fig. \ref{figE3}. From these results, we identified a threshold value of $\theta$ for each $J$ such that $M<1$. These values were utilized to determine the phase boundary between the FM and MD phases, as depicted in Fig. \ref{fig2}\textbf{c} of the main text.

\subsection{Critical twist angle formula for FM-MD transition} \label{sec:theta_c_FM-MD}

We derived the critical twist angle formula using an effective energy functional in the continuum limit, corresponding to the spin Hamiltonian in Eq. \eqref{eq:spinH} of the main text:
\begin{align} \label{eq:spinH3}
    E[\mathbf{n}_l] & = \frac{1}{2}J\sum_{l=t,b}\int d^2\mathbf{r}\left|\nabla \mathbf{n}_l(\mathbf{r})\right|^2+ \frac{A}{a^2}\sum_{l=t,b}\int d^2\mathbf{r}\left(n_{l,z}(\mathbf{r})\right)^2 \nonumber \\
    & + \frac{1}{a^2}\int d^2\mathbf{r}~J_{\perp}\left(\mathbf{r}\right) \mathbf{n}_t(\mathbf{r}) \cdot \mathbf{n}_b(\mathbf{r}),
\end{align}
where $\mathbf{n}_l(\mathbf{r})$ represents the vector field depicting the orientation of spins at position $\mathbf{r}$ in the top ($l=t$) and bottom ($l=b$) layers, respectively. To compute the magnetic energy for the MD phase, we employed several approximations. Firstly, we considered the shape of an AFM patch as a disk with a size of $L$. Secondly, we employed the following function form for the interlayer coupling $J_{\perp}(\mathbf{r})$:
\begin{align} \label{eq:J_perp_r}
    J_\perp(\mathbf{r}) = J_\perp^\textrm{max}(1-|\mathbf{r}|/L),
\end{align}
where $J_\perp^\textrm{max}$ is the maximum value of $J_\perp(\mathbf{r})$ at the center of the patch. Lastly, we assumed $\mathbf{n}_{t,b}(\mathbf{r})=(0,\sin{\phi_{t,b}(\mathbf{r})},\cos{\phi_{t,b}(\mathbf{r})})$ with the following function forms for the phase angles $\phi_{t,b}(\mathbf{r})$:
\begin{align} \label{eq:phi_tb}
    \phi_t(\mathbf{r}) & = +\phi_0(1-|\mathbf{r}|/L), \nonumber \\
    \phi_b(\mathbf{r}) & = -\phi_0(1-|\mathbf{r}|/L),
\end{align}
where $\phi_{t,b}(\mathbf{r})$ reach their maximum values at the center of the disk. Inserting Eqs. \eqref{eq:J_perp_r} and \eqref{eq:phi_tb} into Eq. \eqref{eq:spinH3}, we obtained an energy function as a function of $\phi_0$ as
\begin{align} \label{eq:E_phi1}
    E(\phi_0) & = \pi J \phi_0^2 + \frac{\bar{J}_\perp\pi L^2}{a^2}\bigg(\frac{3\cos\phi_0(\sin\phi_0-\phi_0\cos\phi_0)}{\phi_0^3}\bigg) \nonumber \\
    & \approx \frac{\bar{J}_\perp\pi L^2}{a^2} + \phi_0^2 \bigg(\pi J - \frac{3\bar{J}_\perp\pi L^2}{5a^2} \bigg)+ \mathcal{O}(\phi_0^4),
\end{align}
where $\bar{J}_\perp$ is the average value of defined as $\bar{J}_\perp=\frac{2\pi}{\pi L^2}\int^L_0 drr J_\perp(\mathbf{r})=J_\perp^\textrm{max}/3$. When the sign of the quadratic term changes, the phase transition occurs. Equating the coefficient of the quadratic term to zero, we find a critical AFM patch size $L_c$:
\begin{align} \label{eq:L_c}
    L_c = a\sqrt{\frac{5J}{3\bar{J}_\perp}}.
\end{align}
From the relation $3\pi L^2 = f \frac{\sqrt{3}}{2} L_\textrm{M}^2$, where $f$ is the fraction of spins in the AFM domains and $L_\textrm{M}=\frac{a}{2\sin(\theta/2)}\approx \frac{a}{\theta}$ is the lattice constant of the moiré superlattice ($3$ comes from the fact that there are three AFM patches per moiré unit cell), we find the relation $L=\sqrt{\frac{f\sqrt{3}}{6\pi}}L_\textrm{M} \approx \sqrt{\frac{f\sqrt{3}}{6\pi}}\frac{a}{\theta}$. Using this relation with Eq. \eqref{eq:L_c}, we finally obtain the critical twist angle $\theta_c$ as 
\begin{align} \label{eq:theta_c}
    \theta_c = \sqrt{\frac{\sqrt{3}}{10\pi} \frac{f\bar{J}_\perp}{J}}.
\end{align}

We utilized this formula to fit the numerically calculated FM-MD phase boundary, as presented in Fig. \ref{fig2}\textbf{c} of the main text. The fitting parameter was obtained as $f\bar{J}_\perp \approx 0.56$meV.


\subsection{Determination of MD-MQ phase boundary} \label{sec:MD-MQ boundary}

\begin{figure*}[t!]
    \centering
   \includegraphics[width=.66\textwidth]{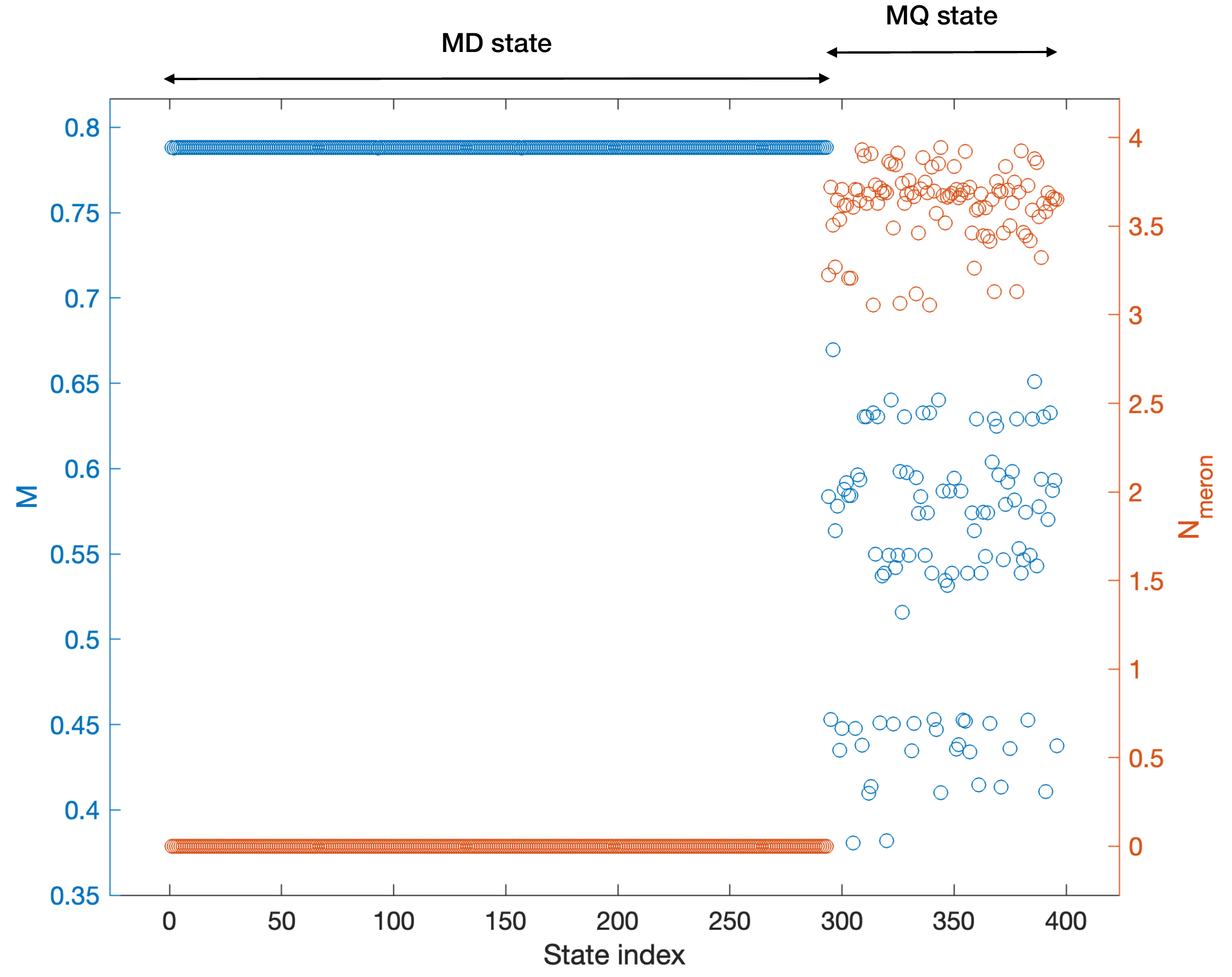}
    \caption[Ext. Data: MD-MQ transition]{\textbf{Left:} Ferromagnetic order parameter ($M$) and \textbf{Right:} total meron number ($N_\textrm{meron}$), as defined in Eqs. \eqref{eq:M} and \eqref{eq:N_meron}, computed from different stable magnetic states. Each state was obtained from a distinct random initial spin configuration. The parameter values of $J=2$meV and $\theta=1.61$\textdegree{} are utilized.}
\label{figE4}
\end{figure*}

We utilized the same relaxed configurations that were employed to determine the FM-MD phase boundary, as depicted in Sec. \ref{sec:FM-MD boundary}. For each spin configuration, we computed the total number of merons from both the top and bottom layers by using the following formula: 
\begin{align} \label{eq:N_meron}
    N_\textrm{meron}=2\times\sum_{l=t, b}\frac{1}{4\pi}\int dx \int dy \left|(\partial_x\mathbf{n}_l\times\partial_y\mathbf{n}_l)\cdot\mathbf{n}_l\right|,
\end{align}
where the integral is performed throughout the moiré unit cell. The absolute value was imposed in order to consider both the number of merons and the number of antimerons, preventing their skyrmion numbers from canceling out with each other. The factor of $2$ comes from the fact that the contribution of a meron or an antimeron is $1/2$. The results for specific parameter values of $J=2$ meV and $\theta=1.61$\textdegree{} are displayed in Fig. \ref{figE4}. We confirmed that the value of $N_\textrm{meron}\approx4$ corresponds to the actual formation of the MQ state. By establishing this correspondence, we computed the total meron number in each relaxed configuration for different values of $J$ and $\theta$. For each $J$, we identified a threshold value of $\theta$ for the appearance of $N_\textrm{meron}\approx4$. These results were utilized to determine the phase boundary between the MD and MQ phases, as depicted in Fig. \ref{fig5}\textbf{a} of the main text.

\subsection{Determination of critical field strengths for MQ-MD transition} \label{sec:critical_field}

\begin{figure}
    \centering
    \includegraphics[width=.66\textwidth]{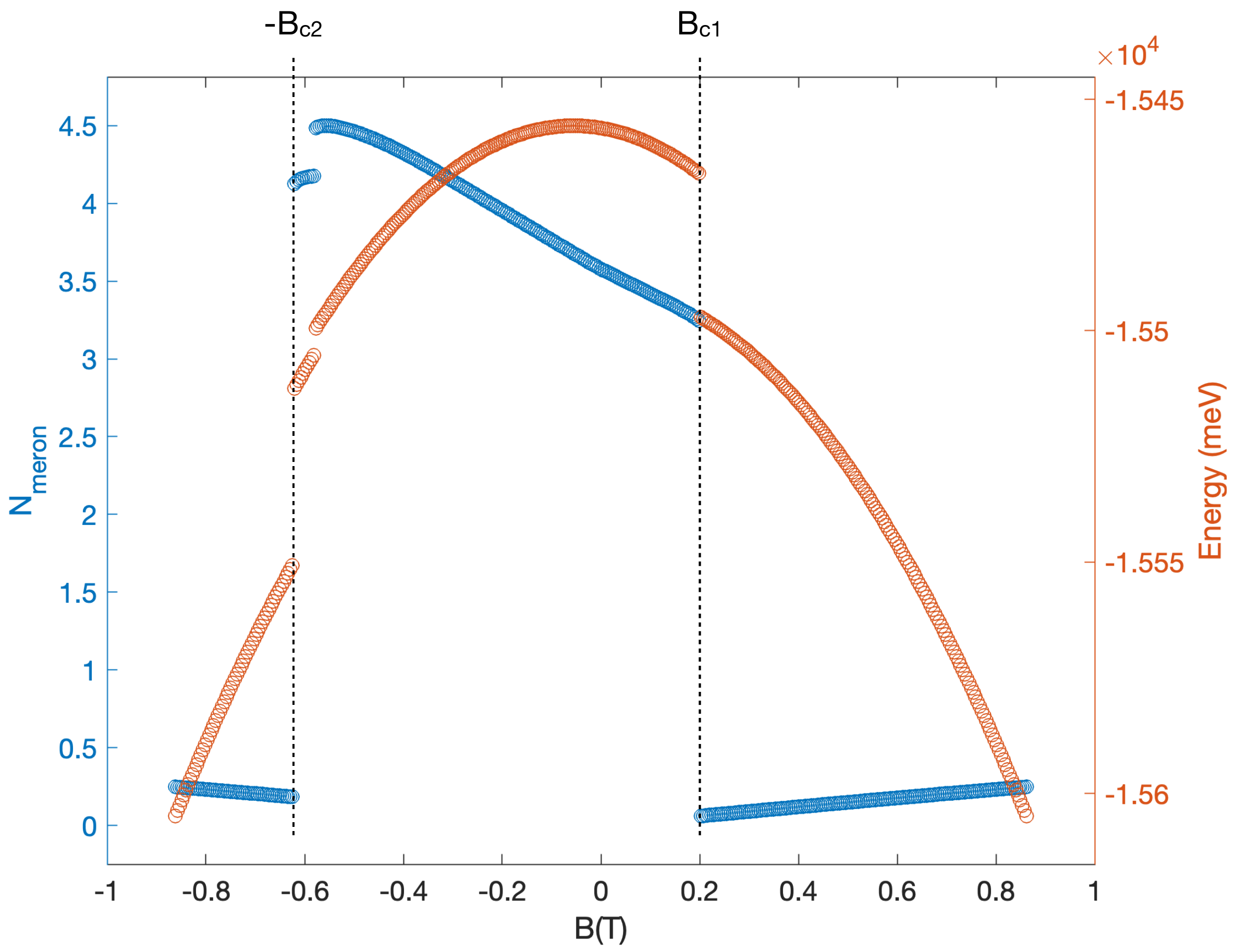}
    \caption[Ext. Data: Critical field for MQ-MD transition]{\textbf{Effect of the Zeeman term, as defined in Eq. \eqref{eq:zeeman} of the main text, on the modification of the MQ state, as depicted in Fig. \ref{fig4}\textbf{a-c}}. The left and right axes correspond to the total meron number ($N_\textrm{meron}$) and magnetic energy, as defined in Eqs. \eqref{eq:N_meron} and \eqref{eq:E_tot}, respectively. The parameter $B$ represents an external magnetic field in the out-of-plane direction. $B_{c1}$ and $B_{c2}$ denote the critical field strengths for the transition of MQ state to the MD state. The parameter values of $J=2$meV and $\theta=1.61$\textdegree{} are utilized. }
    \label{figE5}
\end{figure}

In the first step, we utilized random initial configurations represented as $\{\mathbf{n}_i^{(0)}\}=\{(0,0,1)+0.1*(u_i,v_i,w_i)\}$, where $u_i$, $v_i$, and $w_i$ are random numbers within the range of $[-1,1]$. Normalization was performed to ensure unit length for all $\mathbf{n}_i^{(0)}$. The utilization of these polarized configurations ensured consistency in our analysis by imposing that all MQ states belong to the same class with the same polarity $p=-1/2$ among the meron cores. These initial configurations were relaxed through the iterative optimization method. Through this procedure, we prepared more than 20 different samples of the MQ state in the zero-field condition.

In the subsequent step, we proceeded to compute modified spin configurations from these samples by incorporating the Zeeman term described in Eq. \eqref{eq:zeeman} of the main text. We adjusted the field strength from $B=-0.86T$ to $B=0.86T$ in this term, as depicted in Fig. \ref{fig5}\textbf{b} of the main text. We used $g=2$ and $\mu_B=0.058$meV, and for simplicity, we assumed $S=1$. Each sample was relaxed under the influence of the Zeeman term,  using an iterative optimization method. After the relaxation, we computed the magnetic energy, the total number of merons, and the net magnetization in the out-of-plane direction, using Eqs. \eqref{eq:E_tot} and \eqref{eq:N_meron}, and $M_z = \frac{1}{N}\sum_{i}n_{i,z}$, respectively. We repeated this computation for different field strengths in order to examine the effects of varying magnetic fields. Fig. \ref{figE5} illustrates the results for the magnetic energy and total number of merons. The physical quantities consistently exhibit discontinuous jumps at critical field strengths denoted by $B_{c1}$ and $B_{c2}$. We confirmed the discontinuity corresponds to the actual transition from the MQ to the MD states (we checked that the small jump before $B_{c2}$ corresponds to the reversal of the polarity of the merons). Utilizing this correspondence, we computed the critical field strength for each sample and each parameter of $J$ and $\theta$. These were utilized to determine the critical field strengths for the MQ-MD transition, as depicted in Fig. \ref{fig5}\textbf{c} of the main text.

\subsection{MQ states with different skyrmion numbers}

\begin{figure*}[ht!]
    \centering
    \includegraphics[width=.66\textwidth]{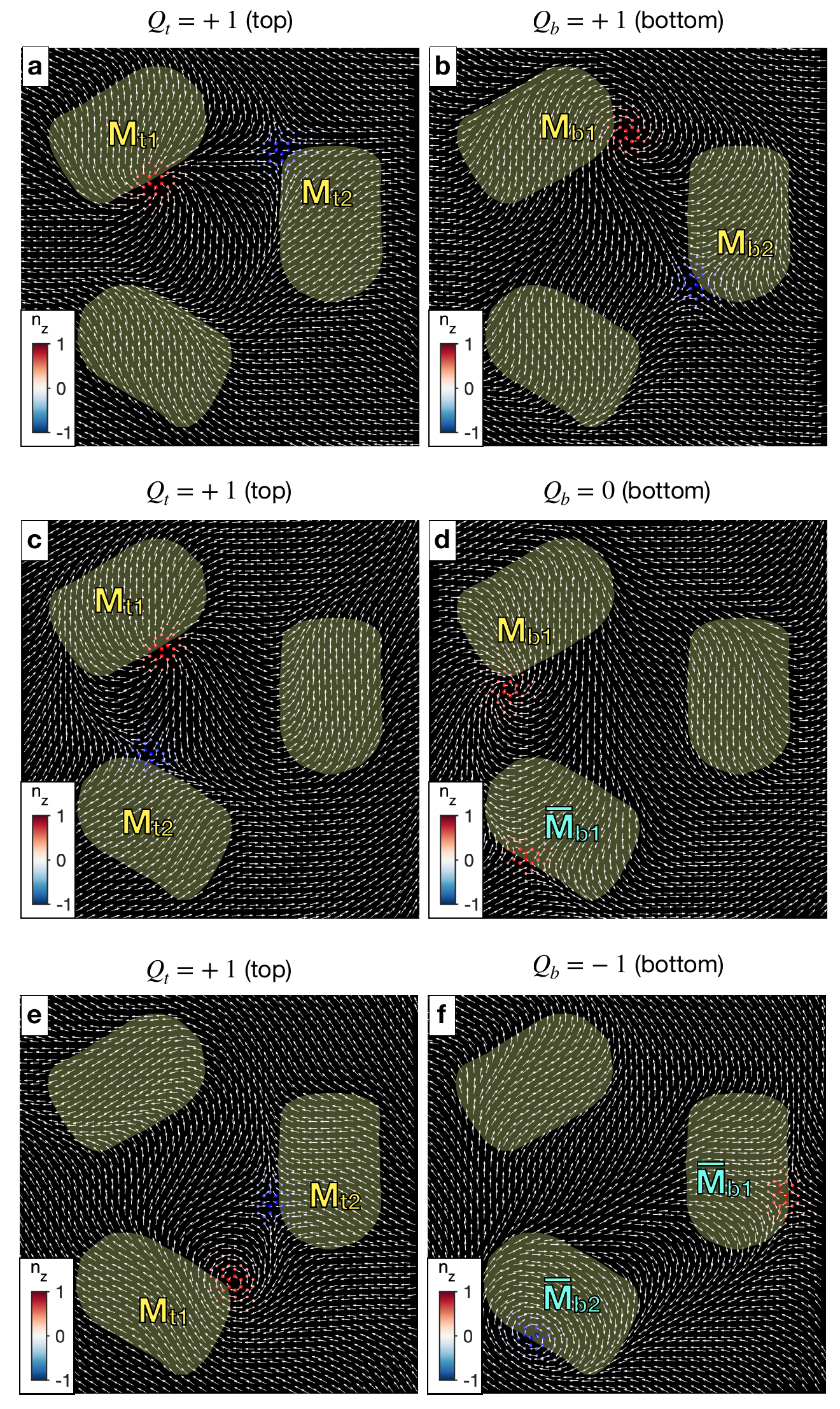}
    \caption[Ext. Data: MQ states with different skyrmion numbers]{Illustrative examples of MQ states showcasing different skyrmion numbers $(Q_t,Q_b)$ in the top and bottom layers, as defined in Eq. \eqref{eq:skyrmion_number_layer}. \textbf{a-b:} $(Q_t,Q_b)=(1,1)$, \textbf{c-d:} $(Q_t,Q_b)=(1,0)$, and \textbf{e-f:} $(Q_t,Q_b)=(1,-1)$. Arrows indicate in-plane components, while the color scale in the markers represents out-of-plane components. Marker sizes are adjusted for better visibility. Shaded areas indicate AFM patches. The parameters $J=2$ meV and $\theta=1.61$\textdegree{} are utilized.}
    \label{figE6}
\end{figure*}

Fig. \ref{figE6} presents illustrative examples of MQ states, which exhibit distinct values of skyrmion numbers $(Q_t, Q_b)$ in the top and bottom layers, respectively, defined as,
\begin{subequations} \label{eq:skyrmion_number_layer}
\begin{align} 
    Q_{t}&=\frac{1}{4\pi}\int dx\int dy(\partial_x\mathbf{n}_t\times\partial_y\mathbf{n}_t)\cdot \mathbf{n}_t, \\
    Q_{b}&=\frac{1}{4\pi}\int dx\int dy(\partial_x\mathbf{n}_b\times\partial_y\mathbf{n}_b)\cdot \mathbf{n}_b.
\end{align}    
\end{subequations}
The first row (\textbf{a-b}) illustrates a pair of two merons in the top layer ($\textrm{M}_\textrm{t1}$--$\textrm{M}_\textrm{t1}$) and a pair of two antimerons ($\overline{\textrm{M}}_\textrm{b1}$--$\overline{\textrm{M}}_\textrm{b2}$) in the bottom layer, corresponding to $(Q_t, Q_b)=(1,-1)$. The second row (\textbf{c-d}) shows a pair of two merons in the top layer ($\textrm{M}_\textrm{t1}$--$\textrm{M}_\textrm{t2}$) and a meron-antimeron pair ($\textrm{M}_\textrm{b}$--$\overline{\textrm{M}}_\textrm{b}$) in the bottom layer, corresponding to $(Q_t, Q_b)=(1,-1)$. Lastly, the third row (\textbf{e-f}) exhibits two pairs of two antimerons in the top layer ($\overline{\textrm{M}}_\textrm{t1}$--$\overline{\textrm{M}}_\textrm{t2}$) and ($\overline{\textrm{M}}_\textrm{b1}$--$\overline{\textrm{M}}_\textrm{b2}$) in the bottom layer, corresponding to $(Q_t, Q_b)=(1,-1)$. The MQ state in Fig. \ref{fig3}\textbf{c-d} of the main text features $(Q_t, Q_b)=(0,0)$.

\section{Supplementary information}

\subsection{Supplementary Video 1} \label{sec:movie1}

Relaxation process for the MQ state, as presented in Fig. \ref{fig4}\textbf{a-c} and \textbf{g-i}. \textbf{First row:} Spin configurations in the top and bottom layers, and relative orientation. Text box in the top right corner displaying the iteration step and the magnetic energy per spin at each step. \textbf{Second row:} Intralayer exchange energy plus single-ion anisotropy energy in the top and bottom layers, and interlayer exchange energy.

\subsection{Supplementary Video 2} \label{sec:movie2}

Relaxation process for the MD state, as presented in Fig. \ref{fig4}\textbf{d-f} and \textbf{j-l}. \textbf{First row:} Spin configurations in the top and bottom layers, and relative orientation. Text box in the top right corner displaying the iteration step and the magnetic energy per spin at each step. \textbf{Second row:} Intralayer exchange energy plus single-ion anisotropy energy in the top and bottom layers, and interlayer exchange energy.



\backmatter

\bmhead{Data availability}
All data generated or analysed during this study are included in this Article. Source data are provided with this paper.

\bmhead{Code availability}
The code written for use in this study is available from the corresponding author upon reasonable request.

\bmhead{Acknowledgements}

K.K. acknowledges support from the Institute for Basic Science in the Republic of Korea under the project IBS-R024-D1. G.G. was supported by the National Research Foundation of Korea (NRF-2022R1C1C2006578). M.J.P. was supported by the National Research Foundation of Korea (NRF) grant funded by the Korea government (MSIT) (RS-2023- 00252085,RS-2023-00218998). Lastly, S.K.K. was supported by Brain Pool Plus Program through the National Research Foundation of Korea funded by the Ministry of Science and ICT (2020H1D3A2A03099291) and National Research Foundation of Korea funded by the Korea Government via the SRC Center for Quantum Coherence in Condensed Matter (RS-2023-00207732).

\end{document}